\documentclass[twoside,twocolumn,9pt]{article}
\usepackage{extsizes}
\usepackage[super,sort&compress,comma]{natbib} 
\usepackage[left=1.5cm, right=1.5cm, top=1.785cm, bottom=2.0cm]{geometry}
\usepackage{balance}
\usepackage{mathptmx}
\usepackage{sectsty}
\usepackage{graphicx} 
\usepackage{lastpage}
\usepackage{multirow}
\usepackage[format=plain,justification=justified,singlelinecheck=false,font={stretch=1.125,small,sf},labelfont=bf,labelsep=space]{caption}
\usepackage{float}
\usepackage{fancyhdr}
\usepackage{fnpos}
\usepackage[english]{babel}
\addto{\captionsenglish}{%
  
}

\usepackage{pdfpages}
\usepackage{array}
\usepackage{droidsans}
\usepackage{booktabs}
\usepackage{charter}
\usepackage[T1]{fontenc}
\usepackage[usenames,dvipsnames]{xcolor}
\usepackage{setspace}
\usepackage[compact]{titlesec}
\usepackage{hyperref}
\usepackage{xfrac}
\usepackage{chemfig}
\usepackage{graphicx}% Include figure files
\usepackage{dcolumn}% Align table columns on decimal point
\usepackage{bm}% bold math
\usepackage{siunitx}
\usepackage{xcolor}

\DeclareSIUnit{\calorie}{cal}

\newcommand{\ket}[1]{\vert #1 \rangle}

\newcommand{\bra}[1]{\langle #1 \vert}

\newcommand{\Tbraket}[3]{\langle #1 \hspace{.10em} \vert \hspace{.10em} #2 \hspace{.10em} \vert \hspace{.10em} #3 \rangle}

\usepackage[version=4]{mhchem}

\newcommand{\simH}{\bar{H}}

\usepackage{epstopdf}%This line makes .eps figures into .pdf - please comment out if not required.

\definecolor{cream}{RGB}{222,217,201}
\begin{document}

\pagestyle{fancy}
\thispagestyle{plain}
\fancypagestyle{plain}{
%%%HEADER%%%
\renewcommand{\headrulewidth}{0pt}
}
%%%END OF HEADER%%%

%%%PAGE SETUP - Please do not change any commands within this section%%%
\makeFNbottom
\makeatletter
\renewcommand\LARGE{\@setfontsize\LARGE{15pt}{17}}
\renewcommand\Large{\@setfontsize\Large{12pt}{14}}
\renewcommand\large{\@setfontsize\large{10pt}{12}}
\renewcommand\footnotesize{\@setfontsize\footnotesize{7pt}{10}}
\renewcommand\scriptsize{\@setfontsize\scriptsize{7pt}{7}}
\makeatother

\renewcommand{\thefootnote}{\fnsymbol{footnote}}
\renewcommand\footnoterule{\vspace*{1pt}% 
\color{cream}\hrule width 3.5in height 0.4pt \color{black} \vspace*{5pt}} 
\setcounter{secnumdepth}{5}

\makeatletter 
\renewcommand\@biblabel[1]{#1}            
\renewcommand\@makefntext[1]% 
{\noindent\makebox[0pt][r]{\@thefnmark\,}#1}
\makeatother 
\renewcommand{\figurename}{\small{Fig.}~}
\sectionfont{\sffamily\Large}
\subsectionfont{\normalsize}
\subsubsectionfont{\bf}
\setstretch{1.125} %In particular, please do not alter this line.
\setlength{\skip\footins}{0.8cm}
\setlength{\footnotesep}{0.25cm}
\setlength{\jot}{10pt}
\titlespacing*{\section}{0pt}{4pt}{4pt}
\titlespacing*{\subsection}{0pt}{15pt}{1pt}
%%%END OF PAGE SETUP%%%

%%%FOOTER%%%
\fancyfoot{}
\fancyfoot[LO,RE]{\vspace{-7.1pt}\includegraphics[height=9pt]{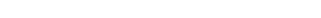}}
\fancyfoot[CO]{\vspace{-7.1pt}\hspace{13.2cm}\includegraphics{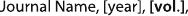}}
\fancyfoot[CE]{\vspace{-7.2pt}\hspace{-14.2cm}\includegraphics{head_foot/RF}}
\fancyfoot[RO]{\footnotesize{\sffamily{1--\pageref{LastPage} ~\textbar  \hspace{2pt}\thepage}}}
\fancyfoot[LE]{\footnotesize{\sffamily{\thepage~\textbar\hspace{3.45cm} 1--\pageref{LastPage}}}}
\fancyhead{}
\renewcommand{\headrulewidth}{0pt} 
\renewcommand{\footrulewidth}{0pt}
\setlength{\arrayrulewidth}{1pt}
\setlength{\columnsep}{6.5mm}
\setlength\bibsep{1pt}
%%%END OF FOOTER%%%

%%%FIGURE SETUP - please do not change any commands within this section%%%
\makeatletter 
\newlength{\figrulesep} 
\setlength{\figrulesep}{0.5\textfloatsep} 

\newcommand{\topfigrule}{\vspace*{-1pt}% 
\noindent{\color{cream}\rule[-\figrulesep]{\columnwidth}{1.5pt}} }

\newcommand{\botfigrule}{\vspace*{-2pt}% 
\noindent{\color{cream}\rule[\figrulesep]{\columnwidth}{1.5pt}} }

\newcommand{\dblfigrule}{\vspace*{-1pt}% 
\noindent{\color{cream}\rule[-\figrulesep]{\textwidth}{1.5pt}} }

\newcommand{\note}[1]{\textcolor{red}{[#1]}}

\makeatother
%%%END OF FIGURE SETUP%%%

%%%TITLE AND AUTHORS%%%
\twocolumn[
  \begin{@twocolumnfalse}
{\includegraphics[height=30pt]{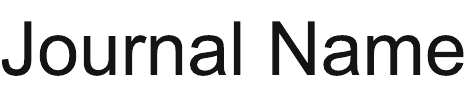}\hfill\raisebox{0pt}[0pt][0pt]{\includegraphics[height=55pt]{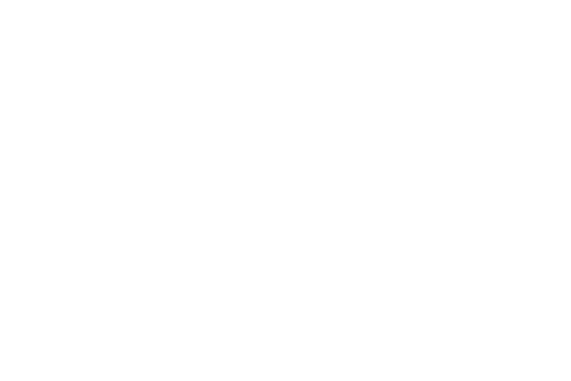}}\\[1ex]
\includegraphics[width=18.5cm]{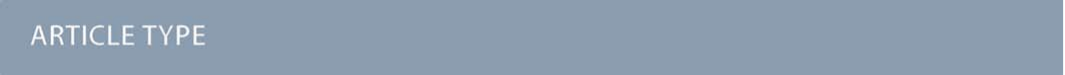}}\par
\vspace{1em}
\sffamily
\begin{tabular}{m{4.5cm} p{13.5cm} }

\includegraphics{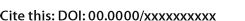} & \noindent\LARGE{Coupled cluster theory for multireference and open-shell molecules} \\
 & \vspace{0.3cm} \\

 & \noindent\large{Sarai Dery Folkestad,$^{\ast}$ Henrik Koch} \\%Author names go here instead of "Full name", etc.

\includegraphics{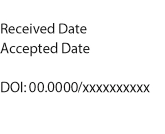} & \noindent\normalsize{We introduce a simple coupled cluster approach for multireference and open-shell systems. In the restricted active space coupled cluster framework, the cluster operator only includes excitations from doubly occupied to virtual orbitals, and this ensures the individual operators commute among each other. The ground and excited states are obtained by diagonalisation of the coupled cluster similarity transformed Hamiltonian in a restricted active space. We demonstrate the framework for the RAS-CCSD model, where single and double excitations are included in the cluster operator, and likewise for a complete active space. We consider systems with multireference character and open-shell systems in a spin-adapted manner. The approach captures significantly more dynamical correlation than without the coupled cluster similarity transformation.} \\

\end{tabular}

 \end{@twocolumnfalse} \vspace{0.6cm}

  ]
%%%END OF TITLE AND AUTHORS%%%

%%%FONT SETUP - please do not change any commands within this section
\renewcommand*\rmdefault{bch}\normalfont\upshape
\rmfamily
\section*{}
\vspace{-1cm}

%%%FOOTNOTES%%%

\footnotetext{\textit{Department of Chemistry and Biomedical Science, Norwegian University of Science and Technology, 7491 Trondheim, Norway; E-mail: sarai.d.folkestad@ntnu.no}}

%Please use \dag to cite the ESI in the main text of the article.
%If you article does not have ESI please remove the the \dag symbol from the title and the footnotetext below.
%\footnotetext{\dag~Supplementary Information available: [details of any supplementary information available should be included here]. See DOI: 00.0000/00000000.}
%additional addresses can be cited as above using the lower-case letters, c, d, e... If all authors are from the same address, no letter is required

\footnotetext{\dag~Supplementary Information available: }

%%%END OF FOOTNOTES%%%

%%%ABSTRACT%%%%

\sffamily{\textbf{ }}\\%The abstrast goes here instead of the text "The abstract should be..."

%%%END OF ABSTRACT%%%%

\rmfamily %Please do not remove this line.

\section*{Introduction}
Coupled cluster theory has been demonstrated to provide an unmatched accuracy in dynamical electron correlation of molecular systems. The unique property is the exponential parametrization, which ensures correct scaling of the correlation energy with system size. Size-extensivity is essential for chemistry, where energy differences are of the utmost importance. 
However, a major challenge with an exponential parametrization arises when the reference state has partial or fractional orbital occupations, such as open-shell molecules or systems where orbitals are nearly degenerate (multireference systems).

Currently, there is no community consensus on how a general multireference coupled cluster theory should be formulated and implemented. Generally, one can divide multireference coupled cluster models into two categories: a pragmatic approach that builds on a single determinant reference and the \textit{genuine}\cite{Lyakh_2011} multireference models, that use a multideterminental reference state. 
The first class of multireference models originates from the work of Oliphant and Adamowicz,\cite{Olipant_1991,Olifant_1992} and has since been used by several others, see e.g., Refs.~\citenum{Piecuch_1993} and \citenum{Kallay_2002}.
This approach is attractive because it exploits the simplicity of single-reference coupled cluster theory, capturing multireference character by including higher-order excitations in a restricted orbital space.  
However, when a qualitative description of the wavefunction has equal (or almost equal) weights for several determinants---as is the case in a dissociation processes---a single reference-based approach is arguably imbalanced. 

A balanced multireference coupled cluster theory must treat each determinant in the zeroth-order wavefunction on an equal footing. 
An appealing strategy is offered by the internally contracted multireference coupled cluster model,\cite{Adam_2025} where a single cluster operator is defined relative to a multideterminental reference.\cite{Mukherjee_1975,Banerjee_1981,Hanauer_2011,Aoto_2016}  
The consequence is non-commuting terms in the cluster operator, leading to prohibitive complexity unless the Baker-Campbell-Hausdorff (BCH) expansion is truncated.
One way to circumvent the complications arising from non-commutativity of the cluster operator is to use normal ordering techniques.\cite{Mukherjee_1975,Lindgren_1978,Nooijen_2001, Gunasekera_2024}
The Fock-space coupled cluster wavefunction\cite{Mukherjee_1975, Lindgren_1978, Pal_1988, Kaldor_1991,Landau_2000} also takes the form of a normal-ordered exponential of a single cluster operator. However, contrary to the internally contracted CC approach, the Fock-space coupled cluster wavefunction is constructed, sector-by-sector in the Fock space, starting from a closed-shell single reference determinant (which may, or may not, have a different number of electrons from the target Fock space sector).

Finally, another class of genuine multireference coupled cluster approaches follows the Jeziorski-Monkhorst\cite{Jeziorski_1981} parametrization, where a separate cluster operator is defined for each determinant in the reference state. The original Jeziorski-Monkhorst approach suffers from redundancy and orbital-invariance problems. However, the parametrization has still been used to formulate both state-universal and state-specific approaches.\cite{Li_2003,Hanrath_2005,Evangelista_2010,Demel_2023}
Irrespective of the parametrization, genuine multireference coupled cluster theory is considerably more complicated than its single-reference counterparts.

Traditionally, and partially due to the complications with multireference coupled cluster theory, the community has resorted to complete active space (CAS) approaches for molecular systems with significant multireference character. These methods include CAS configuration interaction (CASCI), restricted active space configuration interaction (RASCI)\cite{Olsen_1988, helgaker2014}, or their orbital optimized counterparts CASSCF\cite{Roos_1980,Siegbahn_1981} and RASSCF\cite{Malmqvist_1990}. While these approaches capture static correlation, dynamical correlation is not satisfactorily described. The description dynamical correlation can, in principle, be improved by increasing the active space. However, the combinatorial scaling with active space size renders this strategy impractical, even when the density matrix renormalization group (DMRG)\cite{White_1992,White_1999} approximation is used. Hence, dynamical correlation is usually included through second-order perturbation theory (PT2), as in the CASPT2,\cite{andersson_1990} RASPT2,\cite{Malmqvist_2008} and NEVPT2\cite{Angeli_2001} models. A second-order perturbation theory is generally not sufficient for chemical accuracy and not necessarily systematically improvable.
More recently, CAS (and DMRG) approaches have been combined with coupled cluster theory to model dynamical correlation, for instance through energy corrections,\cite{Phung2018} or using Hamiltonian downfolding techniques.\cite{bauman_2025}

%\begin{align*}
%    \ket{\Psi} &= \sum_{\mu}e^{T_\mu}\ket{\mu}C_{\mu}\\
%    \ket{\Psi^{IC}} &= e^{T}\sum_{\mu}\ket{\mu}C_{\mu}\\
%    \ket{\Psi^{NO}} &= \{e^{T}\}\sum_{\mu}\ket{\mu}C_{\mu}\\
%    \ket{\Psi^{LML}} &= e^{T}(1+C^{\text{int}})\ket{0}\\
%    \ket{\Psi^{ML}} &= e^{T}e^{T^{\text{int}}}\ket{0}
%\end{align*}
%+ Fock space CC

A coupled cluster theory for open-shell molecules that enforces the exact spin properties encounter similar complications as those that arise for molecular systems with multireference character. A so-called spin-adapted theory also yields non-commuting terms in the cluster operator that lead to highly complicated equations, even after truncating the BCH-expansion.\cite{Janssen_1991, Jeziorski_1995, Li_1994, Neogrady_1994, Herrmann_2020, Herrmann_2022}
One highly successful approach for open-shell states that formally have a single-determinant reference, is the unrestricted, or spin-orbital coupled cluster theory. Unrestricted coupled cluster theory is simple in both formulation and implementation. However, the unrestricted coupled cluster wavefunction is not an eigenfunction of the ${S}^2$ operator, and the approach is spin-contaminated. While the spin contamination of the unrestricted coupled cluster ground state\cite{Stanton_1994} is small, even for systems where the unrestricted Hartree--Fock spin contamination is significant, the equation-of-motion framework for excited states\citep{Sekino_1984,Geertsen_1989,Krylov_2008,Bartlett_2012} is highly affected by spin-contamination unless a restricted open-shell Hartree--Fock reference is used.\cite{Krylov_2017} 

Multireference and open-shell systems are not the only complications for coupled cluster theory. In general, the theory gives unphysical results close to state degeneracies. Due to the non-Hermitian structure, the theory encounters complex eigenvalues and regions where the matrices are non-invertible. Similarity-constrained coupled cluster theory\cite{Kjonstad_2017,Kjonstad_2019} has been developed to address this problem, where a similarity transformation is used to enforce real eigenvalues. This has made the use of coupled cluster theory in non-adiabatic dynamics simulations possible.\cite{Kjonstad_2024}

Recently, the inability of coupled cluster theory to describe the geometric phase effect has become known.\cite{Kjonstad_2025_geometric} Due to the intermediate normalization of the wavefunctions, artifacts may appear close to regions where the reference contribution to the wavefunction is zero. Generalized coupled cluster theory\cite{Rossi_2025,Stoll_2026} has solved this problem by enforcing convexity in the coupled cluster equations, and the subsequent diagonalisation of the similarity transformed Hamiltonian ensures the correct description of the geometric phase. A correct description of the geometric phase should be a target as we design new electronic structure models.

\begin{figure}
    \centering
    \includegraphics[width=0.3\linewidth]{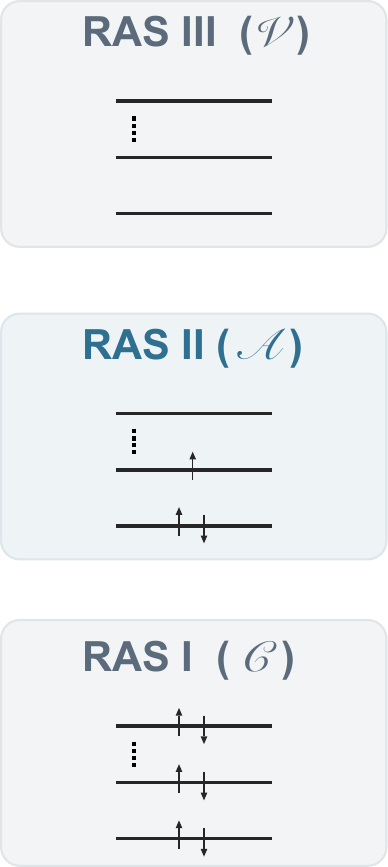}
    \caption{Definition of the RAS orbital partitioning ($\mathcal{O} = \mathcal{C} \cup  \mathcal{A} \cup  \mathcal{V},$). In this work we allow up to two holes and particles in RAS I ($\mathcal{C}$) and III ($\mathcal{V}$), respectively.}
    \label{fig:ras}
\end{figure}

In this paper, we present a balanced multireference coupled cluster framework that is, in many ways, simpler than the previously explored formulations of the theory. In our approach, the cluster operator is restricted to excite electrons from doubly occupied orbitals into the virtual orbitals, avoiding the issue of non-commutativity. The reference state can be either restricted Hartree--Fock, CASSCF, or state-averaged CASSCF, depending on the problem at hand. Finally, the similarity-transformed Hamiltonian is diagonalized in a restricted active space basis. We demonstrate the capabilities of the framework by considering the ethylene molecule and its cation, showing that the model provides a balanced description of multireference and open-shell character, and that the model is both size-extensive and size-intensive for a fixed active space.  

\begin{figure*}[tbh!]
\centering
  \includegraphics[width=0.3\linewidth]{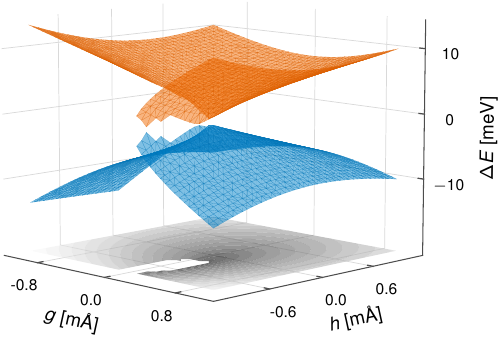}  \includegraphics[width=0.3\linewidth]{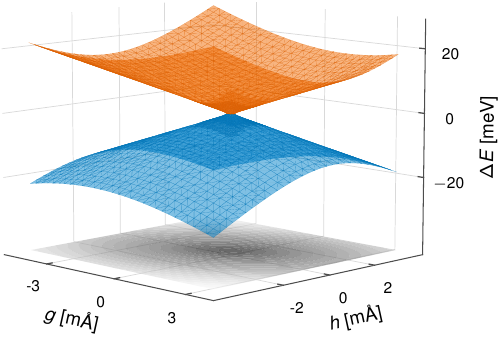}
\includegraphics[width=0.3\linewidth]{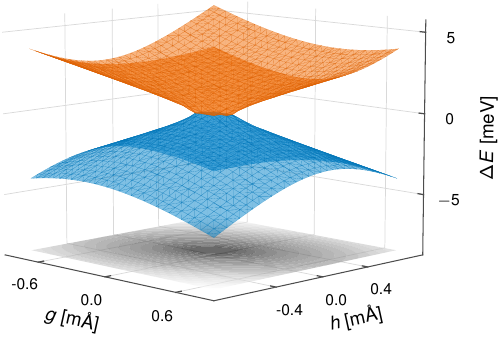}
  \caption{CCSD (left) and RAS-CCSD (middle and right) for the ground state conical intersection of ethylene/cc-pVDZ, using a restricted Hartree--Fock reference. Energies are plotted relative to the average energy of the two states $\Delta E = E_i - \frac{1}{2}(E_1 + E_2)$. 
  Close to the conical intersection, the CCSD equations fail to converge.  While the RAS-CCSD calculations have no convergence issues, the approach is non-Hermitian and yields complex eigenvalues in a region around the intersection point.  \label{fig:ethylene_2d}}
\end{figure*}
\begin{figure*}
    \centering
    \includegraphics[width=0.7\linewidth]{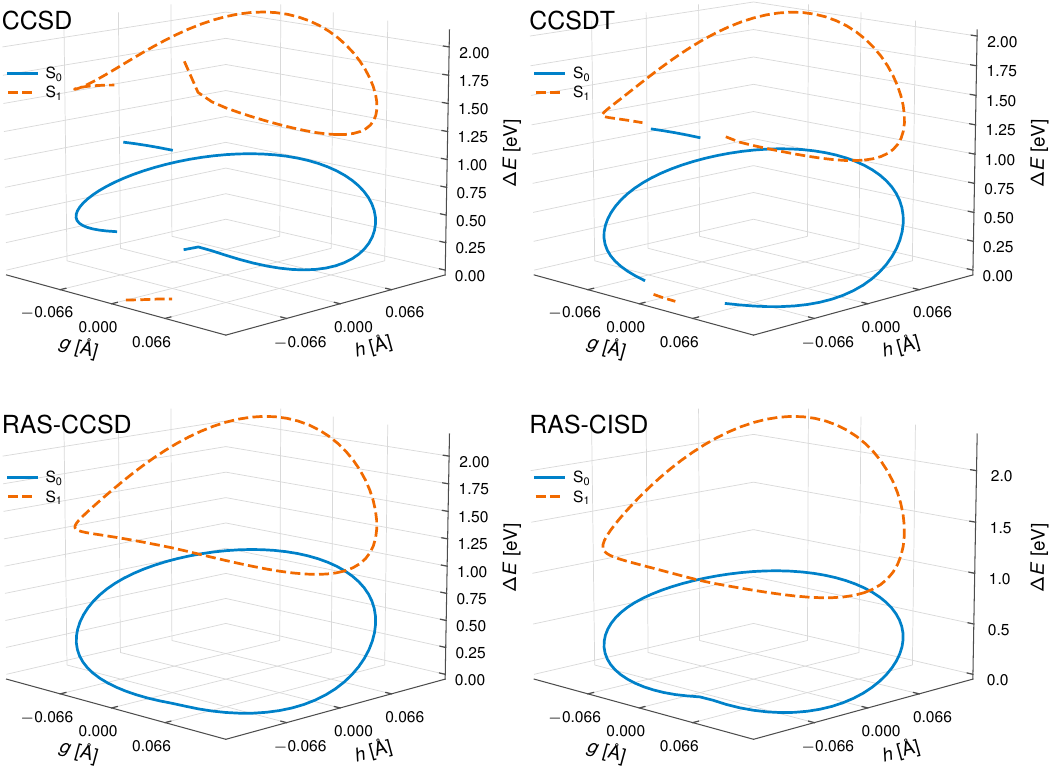}
    \caption{Traversing in a loop of radius $r=\SI{0.25}{\angstrom}$ around the approximate CCSD/cc-pVDZ conical intersection, in the $gh$-plane provided in the Supporting information of Ref.~\citenum{Rossi_2025}. Restricted Hartree--Fock orbitals are used in all calculations. The CCSD and CCSDT models show convergence failures, multiple solutions, and regions of negative excitation energies. The RAS-CCSD and RAS-CISD calculations encounter no problems with convergence, traversing a loop around the conical intersection.}
    \label{fig:ethylene_loop}
\end{figure*}

\section*{Theory}
Coupled cluster theory incorporates dynamical correlation compactly through its exponential parametrization of the wavefunction,
\begin{align}
 \ket{\Psi_0} =  \exp{(T)}\ket{0} = \ket{0} + T \ket{0} + \tfrac{1}{2} T^2 \ket{0} +\ldots.
\end{align}
Disconnected cluster excitations ensure size-extensivity, that is, that the correlation energy scales correctly with system size. The cluster operator is expressed as a sum of excitations operators $\tau_\mu$ with an associated amplitude $t_\mu$,
\begin{align}
 T = \sum_\mu t_\mu \tau_\mu,
\end{align}
where index $\mu$ labels the excitations.
The cluster amplitudes are determined by solving the projected coupled cluster equations:
\begin{align}
    \Omega_\mu = \Tbraket{\mu}{\exp(-T)H\exp(T)}{0} = 0.\label{eq:cc}
\end{align}
In equation-of-motion coupled cluster theory, the similarity transformed Hamiltonian, $\simH = \exp(-T)H\exp(T)$, is diagonalized to give both ground and excited states of the molecular system. 
Solving the coupled cluster equations, i.e., determining $\boldsymbol{t}$, can be viewed as a preparation step before the truncated configuration interaction diagonalisation. The similarity-transformation transforms the problem into a non-Hermitian eigenvalue problem that yields a superior description of dynamical correlation.
When the eqs.~\eqref{eq:cc} are solved, the equation-of-motion Hamiltonian matrix takes the form
\begin{align}
  \boldsymbol{\simH}^{\text{CC}} =  \begin{pmatrix}
    E_{\text{CC}} & \boldsymbol{\eta}^T \\
    \boldsymbol{0} & \boldsymbol{A} + E_{\text{CC}}\boldsymbol{I}
  \end{pmatrix},
\end{align}
where the coupled cluster energy is $E_{\text{CC}} =  \Tbraket{0}{\simH}{0}$, $\eta_\mu = \Tbraket{\mu}{\simH}{0}$, and the coupled cluster Jacobian
is given by
\begin{align}
    A_{\mu\nu} =  \Tbraket{\mu}{[\simH, \tau_\nu]}{0}.
\end{align}
The lowest eigenvalue is usually the coupled cluster energy of the ground state. When $T$ includes all possible excitations, the full coupled cluster model is obtained, which in principle recovers the full configuration interaction (FCI) energies.
%The exponential parametrization causes, however, a few severe failings of the theory, some of which are not fixed at the full coupled cluster limit.

In this paper, we present a strategy to address the increase in complexity of the coupled cluster equations when one wishes to either achieve spin-adaptation for open-shell molecules or consider a reference wavefunction consisting of several determinants. The strategy is pragmatic: we only include excitations from doubly occupied orbitals to virtual orbitals in $T$, and introduce correlation for partially filled (active) orbitals through restricted active space (RAS) diagonalisation of $\simH$. The RAS orbital partitioning is shown in Figure \ref{fig:ras}, where the orbitals are divided into three sets (RAS I--III),
\begin{align}
    \mathcal{O} = \mathcal{C} \cup  \mathcal{A} \cup  \mathcal{V},
\end{align}
consisting of the doubly occupied core orbitals (RAS I, $\mathcal{C}$), the active orbitals (RAS II, $\mathcal{A}$), and the virtual orbitals (RAS III, $\mathcal{V}$). 
The RAS determinant basis is obtained by restricting the maximal number of holes and particles in RAS I and III, respectively. 

The first step of the RAS-CC scheme is a standard (closed-shell spin-adapted) coupled cluster calculation for the RAS I and RAS III orbitals, i.e., eqs. \eqref{eq:cc}
with $T$ restricted to excite from orbitals in $\mathcal{C}$ to orbitals in $\mathcal{V}$. For the RAS-CCSD model, which we consider here, the cluster operator is given by
\begin{align}
    T = \sum_{a\in\mathcal{V}}\sum_{i\in\mathcal{C}} t^{a}_i E_{ai} + \frac{1}{2}\sum_{a,b\in\mathcal{V}}\sum_{i,j\in\mathcal{C}} t^{ab}_{ij} E_{ai}E_{bj}.
\end{align}
The electrons occupying the active space contribute only to the coupled cluster equations through contributions to the generalized Fock matrix,
\begin{align}
\begin{split}
        F_{pq} &=  h_{pq} + \sum_{i\in\mathcal{C}}(2g_{pqii} - g_{piiq}) \\
        &+ \sum_{v,w\in\mathcal{A}}D_{vw}(g_{pqvw} - \tfrac{1}{2}g_{pwvq}),\;\; p,q\in \mathcal{C}\cup\mathcal{V},
\end{split}
\end{align}
and we note that the molecular orbital density matrix for the RAS II orbitals ($D_{vw},\;vw\in\mathcal{A}$) can be obtained from a restricted Hartree--Fock calculation (open- or closed-shell), a CASSCF calculation, or a state-averaged CASSCF calculation. We will show examples of all these reference wavefunctions below. 

After solving the standard coupled cluster equation, $\simH$ is diagonalized in a RAS determinant basis ($\boldsymbol{\simH}^{\text{RAS-CC}}$). We note that the similarity-transformed Hamiltonian is non-Hermitian and that properties (other than energies) are generally obtained only after both left and right eigenvectors are determined:
\begin{align}
  &\boldsymbol{\simH}^{\text{RAS-CC}} \boldsymbol{R}_k = E_k \boldsymbol{R}_k,\quad
   \boldsymbol{L}_k^T\boldsymbol{\simH}^{\text{RAS-CC}} = E_k \boldsymbol{L}_k^T,\quad
  \boldsymbol{L}_k^T \boldsymbol{R}_l = \delta_{kl}.
\end{align}
The RAS-CCSD Hamiltonian matrix elements takes the form
\begin{align}
\begin{split}
    {\simH}^{\text{RAS-CCSD}}_{\mu\nu} &= \Tbraket{\mu}{e^{-T}H e^{T}}{\nu}\\
    &= \Tbraket{\mu}{\tilde{H} + [\tilde{H},T_2] + \frac{1}{2}[[\tilde{H},T_2],T_2]}{\nu} \\
    &= {\tilde{H}}_{\mu\nu} + \Tbraket{\mu}{[\tilde{H},T_2] + \frac{1}{2}[[\tilde{H},T_2],T_2]}{\nu},
\end{split}
\end{align}
where $\tilde{H}$ is the $T_1$-transformed Hamiltonian,\cite{helgaker2014} and $\mu$ and $\nu$ now label determinants in the restricted active space. Note that for $T_1 = 0$, ${\tilde{H}}_{\mu\nu}$ is the RAS-CI Hamiltonian matrix.
If we compare the diagonalisation step of the RAS-CCSD approach to a RAS-CISD implementation, we have two additional considerations: the loss of symmetry in the integrals and additional terms arising from the commutators with $T_2$. Though the additional $T_2$-dependent terms imply evaluating three- and four-body density matrix elements, the elements that enter the equations never have more then four general orbital indices, and by exploiting the restriction of maximally two holes and particles in RAS I and III, respectively, the computational scaling does not exceed that of the underlying non-Hermitian (when $T_1\neq 0$) RAS-CISD problem. 

An important feature of the RAS-CC approach, is that the active orbital space (RAS II, $\mathcal{A}$) is kept minimal, letting the $T$ operator include as much dynamical correlation as possible. In this way, we avoid the ambiguous active space selection in CAS based approaches; the active space should always be the minimal space to capture the chemical process. For instance, a single bond dissociation generally requires the use of CAS(2,2), a double bond dissociation requires the use of CAS(4,4), and a doublet state will use a CAS(1,1), unless orbital degeneracy (due to molecular point-group symmetry) must be taken into account. 

\section*{Results and Discussion}
The RAS-CCSD code is implemented in a development version of the e$^{T}$ program.\cite{et-v1.0-jcp-2020,et-v2.0-jcp-2026}
When a RHF or ROHF reference is used, it is directly in e$^{T}$. However, when a CASSCF or SA-CASSCF reference state is used, the orbitals and active orbital density ($D_{vw}$, $v,w\in\mathcal{A}$) is obtained from a separate PySCF\cite{sun2018pyscf,sun2020recent,sun2026python} calculation. A cc-pVDZ basis is used throughout. Details of the active spaces and reference state types of the calculations are given in Table \ref{tab:details}.

\begin{table}[]
    \centering
    \caption{Details of the active space, and reference state type for all calculations in  this paper.}
    \begin{tabular}{l c c c}
    \toprule
    Calculation & CAS space & Reference type \\
    \midrule 
    2D scan (Figure \ref{fig:ethylene_2d}) & CAS(2,2) & RHF \\
    Loop scan (Figure \ref{fig:ethylene_loop}) & CAS(2,2) & RHF \\
    Ethylene torsion (Figure \ref{fig:ethylene_torsion}) & CAS(2,2) & CASSCF(2,2)/RHF \\
    Ethylene stretch (Figure \ref{fig:ethylene_stretch}) & CAS(4,4) & CASSCF(4,4) \\
    Ethylene cation $D_{2h}$ (Table \ref{tab:ethylene_cation_d2h}) & CAS(1,1) & ROHF \\
    Ethylene cation $D_{2d}$ (Table \ref{tab:ethylene_cation_d2d}) & CAS(1,2) & SA-CASSCF(1,2) \\
    \bottomrule
    \end{tabular}
    \label{tab:details}
\end{table}

We first demonstrate the flexibility of the RAS-CCSD framework for ethylene.
In Figure \ref{fig:ethylene_2d}, the ground state conical intersection in ethylene is shown for both CCSD and RAS-CCSD. The failure of coupled cluster theory to converge close to a ground state degeneracy is evident, and we note that the calculation sequence and restarting strategy will affect the CCSD calculations, due to the presence of multiple solutions. Details of the procedure is given in the \textit{Supporting Information}$^\dag$, together with the coordinates that define the scan. Since the Jacobian is positive definite, the RAS-CCSD model does not encounter convergence issues, nor issues with multiple solutions. However, the RAS-CC Hamiltonian is non-Hermitian and complex energies will occur close to the conical intersection.
The region of complex energies has an approximate radius of $\SI{0.115}{\milli\angstrom}$. The issue of complex energies can be resolved through a strategy similar to the similarity constrained coupled cluster (SCC)\cite{Kjonstad_2017,Kjonstad_2019} model can be taken to eliminate the complex eigenvalues. 

The failure of coupled cluster theory around ground state conical intersections can also be seen by traversing a loop around the intersection, as is shown in Figure \ref{fig:ethylene_loop} for ethylene. The calculation sequence is described in the \textit{Supporting Information}$^\dag$ and restricted Hartree--Fock orbitals are used. Both the CCSD and CCSDT calculations display convergence failures, multiple solutions (discontinuities), and regions of negative excitation energies.
This is the failure to capture the geometric phase effect.\cite{Kjonstad_2025_geometric}
The RAS-CCSD and RAS-CISD models display neither convergence problems nor multiple solutions. We note that the RAS-CISD model has a cusp, caused by discontinuities in the orbital energy derivatives evident from the HOMO and LUMO orbital energies, see the \textit{Supporting Information}$^\dag$.

\begin{figure}[tbh!]
\centering
\includegraphics[width=0.8\linewidth]{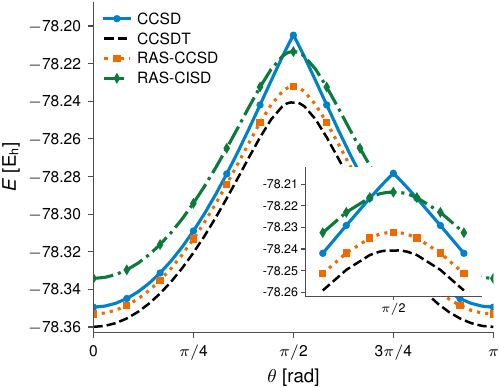}
  \caption{Ethylene/cc-pVDZ C-C bond torsion (unrelaxed) with CCSD, CCSDT, RAS-CCSD and RAS-CISD. The CCSD and CCSDT calculations, and the CASSCF(2,2) reference calculations, are calculated by restarting from the previous point (increasing $\theta$).\label{fig:ethylene_torsion}}
\end{figure}
In Figure \ref{fig:ethylene_torsion},  we consider the ethylene torsion between the planar ($D_{2h}$) and twisted ($D_{2d}$) geometries. The RAS-based approaches are compared to CCSD and CCSDT. CASSCF(2,2) orbitals are used in the RAS-CCSD and RAS-CISD calculations. The CCSD calculations (and to a lesser extent CCSDT) shows a cusp at the $D_{2d}$ geometry, due to orbital degeneracies. In Table \ref{tab:ethylene:torsion_energy_diff}, the energy relative to CCSDT is given for RAS-CCSD, RAS-CISD and CCSD. The breakdown of the CCSD calculation close to the $D_{2d}$ geometry is evident, while the RAS-CCSD calculation has similar accuracy at every angle. Again, RAS-CISD is seen to lack dynamical correlation.

\begin{figure}[tbh!]
\centering
  \includegraphics[width=0.8\linewidth]{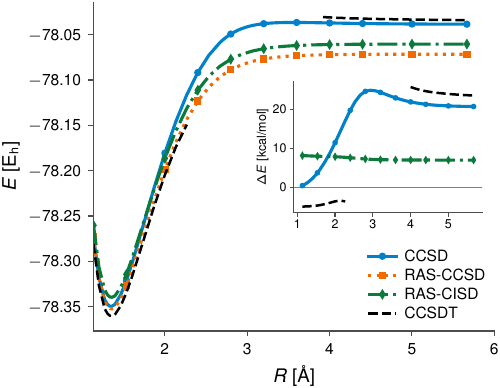}
  \caption{Ethylene/cc-pVDZ C-C bond dissociation (unrelaxed) with CCSD, CCSDT, RAS-CCSD, and RAS, RAS-CISD. The CCSD, CCSDT, and the CASSCF(4,4) reference calculations, are calculated by restarting from the previous point. The inset shows the energies relative to RAS-CCSD, $\Delta E = E_{\text{RAS-CCSD}} - E_{\text{X}}$.\label{fig:ethylene_stretch}}
\end{figure}

In Figure \ref{fig:ethylene_stretch}, the C-C bond of ethylene ($D_{2h}$) is stretched, using CASSCF(4,4) orbitals in the RAS-CCSD and RAS-CISD calculations. The RAS-approaches are compared to CCSD and CCSDT.
The inset shows the energies relative to RAS-CCSD, $\Delta E = E_{\text{RAS-CCSD}} - E_{\text{X}}$.
The CCSDT model fails to converge for the intermediate bond lengths, while CCSD goes through a weak maximum. Both methods converge to the wrong state for larger bond distances. On the other hand, the RAS-models correctly describe the dissociation process and RAS-CCSD recovers more dynamical correlation than RAS-CISD. In the \textit{Supporting Information}$^\dag$, we show that a similar picture is obtained with RAS-CCSD for the dissociation of N$_2$ and for the C--N dissociation of CH$_3$CN, both of which are triple bond stretches.

\begin{table}
  \centering
  \caption{Deviation from CCSDT along the ethylene/cc-pVDZ torsion,
    $\Delta E = E_\mathrm{model} - E_\mathrm{CCSDT}$ in $\si{\kilo\calorie\per\mol}$ at $\theta = 0$ and $\theta = \tfrac{\pi}{2}$.
    The RAS methods use a CAS(2,2) active space, we report results using both RHF and CASSCF references for the RAS. \label{tab:ethylene:torsion_energy_diff}}
  \label{tab:torsion-errors}
  \begin{tabular}{llccc}
    \hline
    Model  & Reference & $\theta = 0\; (D_{2h})$ & $\theta = \sfrac{\pi}{2}\; (D_{2d})$ \\
    \hline
    \multirow{2}{*}{RAS-CCSD} & CASSCF &  ${4.29}$  & ${5.32}$ \\
     & RHF   &  $5.00$  & $6.19$ \\
    \multirow{2}{*}{RAS-CISD} & CASSCF &  ${16.16}$ & ${17.00}$ \\
    & RHF &  $17.77$ & $20.29$ \\
    CCSD     & RHF   &  ${6.58}$  & ${22.48}$ \\
    \hline
  \end{tabular}
\end{table}
\begin{table*}[htbp]
  \centering
  \caption{Ethylene cation/cc-pVDZ ground state total energy $E_0$ (in $E_h$)
    and excitation energies of the lowest doublet states (in eV) at the
    planar ($D_{2h}$) geometry.  An ROHF reference is used in the RAS-CCSD, RAS-CISD, and RAS-CISDT calculations. The EOM-IP-CCSD and EOM-IP-CCSDT use the closed-shell (neutral) Hartree--Fock reference.
    We also report the percentage of doubles excitation character for the
    EOM-IP-CCSDT calculation ($|R_2|^2$), and compare to ROHF-UCCSD and ROHF-UCCSDT calculations performed with QChem.\cite{epifanovsky2021software}}
  \label{tab:ethylene_cation_d2h}
  \begin{tabular}{lcccccccccc}
    \hline
    & EOM-IP-CCSD & EOM-IP-CCSDT  & \% doubles & RAS-CCSD & RAS-CISD & RAS-CISDT & ROHF-UCCSD  ($\langle S^2\rangle$)& ROHF-UCCSDT \\
    \hline
    $E_0$           & -77.9671 & -77.9777 & 3.5  & -77.9714 & -77.9550 & -77.9612 & -77.9714 (0.7501)  & -77.9777\\
    $\omega_1$      & 2.497    & 2.492    & 4.8  & 2.650    & 6.222    &   2.671  & 2.561 & --\\
    $\omega_2$      & 4.203    & 4.142    & 5.4  & 4.256    & 7.855    &   4.319  & 4.215 & --\\
    $\omega_3$      & 5.638    & 5.526    & 12.6 & 5.765    & 9.438    &   5.663  & 5.651 & --\\
    $\omega_4$      & 8.954    & 6.982    & 95.8 & 6.706    & 11.277   &   7.893  & 6.872 & --\\
    \hline
  \end{tabular}
\end{table*}

\begin{table*}
\caption{Ethylene cation/cc-pVDZ ground state total energy $E_0$ (in $E_h$)
    and excitation energies of the lowest doublet states (in eV) at the
    twisted ($D_{2d}$) geometry. The RAS-CCSD and RAS-CISD calculations use a state averaged CASSCF(1,2) as a reference, and the results are compared to ROHF-UCCSD and ROHF-UCCSDT calculations performed with QChem.\cite{epifanovsky2021software}
  \label{tab:ethylene_cation_d2d}}
  \begin{tabular}{lccccc}
    \hline
    & RAS-CCSD & RAS-CISD & RAS-CISDT & ROHF-UCCSD ($\langle S^2\rangle$) & ROHF-UCCSDT \\
    \hline
    $E_0$& -77.9223 & -77.9037 & -77.9159 &-77.9236 (0.7502) & -77.9311\\
    $\omega_1$      & 0.000    & 0.000    & 0.000   & 0.064 & --\\
    $\omega_2$      & 6.568    & 9.611    & 6.251   & 5.890 & --\\
    $\omega_3$      & 7.123    & 10.238   & 6.845   & 6.379 & --\\
    $\omega_4$      & 7.123    & 10.238   & 6.845   & 6.411 & --\\
    \hline
  \end{tabular}
\end{table*}

\begin{table}[]
    \centering
    \caption{Size extensivity errors in RAS-CISD and RAS-CCSD for the molecular systems in Figure \ref{fig:SEsystems}. We characterize the type of size-extensivity based on whether 1) RAS II is divided between subsystems $A$ and $B$ and 2) RAS II is located on subsystem $A$.}
    \begin{tabular}{c c c c}
    \hline
    & \multicolumn{2}{c}{$\Delta E \;[\si{\kilo\calorie\per\mol}]$}&\\
    \cmidrule(lr){2-3}
    Complex & RAS-CISD  & RAS-CCSD & Case\\
    \hline
    \ce{2CH2} & ${6.25}$ & ${1.98}$ & 1\\
    \ce{2CH2} + \ce{2He} & ${11.53}$ & ${1.98}$ & 1, 2\\
    \hline
    \ce{2CH3} & ${9.55}$ & ${0.88}$ & 1\\
    \ce{H2O}+\ce{2He} & ${3.98}$ & ${0.00}$ & 2\\
    \hline
    \end{tabular}
    \label{tab:SEnumbers}
\end{table}

Finally, we consider the ethylene cation in the planar ($D_{2h}$) and twisted ($D_{2d}$) geometries. The ground state energy and the first four doublet excitation energies are given in Tables \ref{tab:ethylene_cation_d2h} and \ref{tab:ethylene_cation_d2d}. In the $D_{2h}$ geometry, we compare the RAS-approaches to EOM-IP-CCSD, EOM-IP-CCSDT, and unrestricted (spin-orbital) CCSD and CCSDT (ROHF-UCCSD and ROHF-UCCSDT) calculations performed with the QChem software;\cite{epifanovsky2021software} all open-shell calculations use ROHF orbitals, while the EOM-IP calculations use closed-shell RHF for the neutral molecule as a reference. The RAS-CCSD and ROHF-UCCSD models agree to all reported digits on the ground state energy of planar ethylene, and the difference compared to EOM-IP-CCSDT, which is assumed to be the most accurate model of the comparison, for both models is $\sim \SI{6}{\milli\hartree}$. The RAS-CCSD model yields excitation energies that differ around $\SI{0.1}{\eV}$ to the UCCSD excitation energies, and $0.2$--$\SI{0.3}{\eV}$ compared to EOM-IP-CCSDT, but shows a great improvement compared to RAS-CISD, which is off by several $\si{\eV}$. The EOM-IP-CCSD model performs very well for the direct ionizations, but does not capture the shake-up ($\omega_4$), which is identified from the double excitation character of the EOM-IP-CCSDT states. 
The twisted geometry arguably provides a more complicated test case.
In the $D_{2d}$ geometry, we compare the RAS-approaches to the unrestricted (spin-orbital) CCSD and CCSDT (ROHF-UCCSD and ROHF-UCCSDT) calculations performed with the QChem software. The ROHF reference used in the unrestricted coupled cluster calculations is symmetry broken. The RAS models use a state averaged CASSCF(1,2) reference, and all RAS models can reproduce a degenerate ground state, and the degeneracy of $\omega_3$ and $\omega_4$, which the UCCSD model cannot achieve, building on symmetry broken orbitals. We emphasize that the RAS-CCSD approach, with the appropriate choice of reference state, is adapted both with respect to spatial symmetry and spin symmetry. 
From the planar cation calculations, we see that, compared to EOM-IP-CCSDT, the RAS-CCSD and RAS-CISDT models produce excitation energies of similar quality (though the RAS-CCSD ground state is significantly better). In the twisted geometry, the models differ more in the produced excitation energies, the RAS-CCSD model giving higher excitation energies. This could indicate that the RAS-CCSD excitation energies are somewhat overestimated, though it is not clear which model is better. 

\begin{figure}
    \centering
    \includegraphics[width=\linewidth]{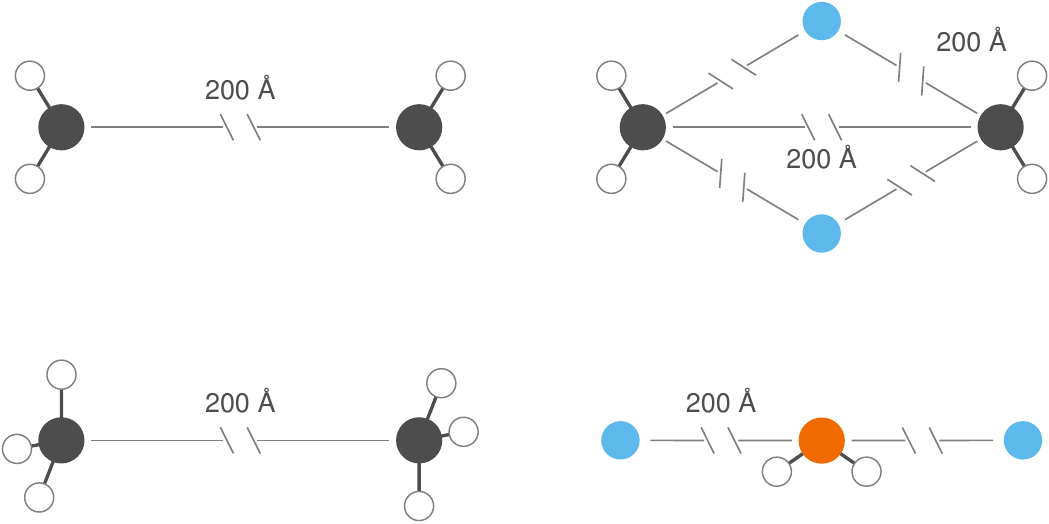}
    \caption{Molecular used in numerical size-extensivity tests: \ce{2CH2} (top left), \ce{2CH2} + \ce{2He} (top right), \ce{2CH3} (bottom left), and \ce{H2O}+\ce{2He} (bottom right)}
    \label{fig:SEsystems}
\end{figure}

\begin{figure*}[tbh!]
\begin{align}
    \simH = \begin{pmatrix}
        E^{A}_{\text{HF}} + E^B_{\text{CC}} & \Tbraket{\text{HF}_A}{H_A}{\nu_A} & \Tbraket{\text{HF}_B}{\simH_B}{\nu_B} & 0\\
        \Tbraket{\mu_A}{H_A}{\text{HF}_A} & H^{A}_{\mu_A\nu_A} + \delta_{\mu_A\nu_A}E^B_{\text{CC}} &  0 & \delta_{\nu_A\mu_A}\Tbraket{\text{HF}_B}{\simH_B}{\nu_B}\\
        \boldsymbol{0} & 0 &\simH^{B}_{\mu_B\nu_B} +\delta_{\mu_B\nu_B} E_{\text{HF}}^{A} &\delta_{\nu_B\mu_B}\Tbraket{\text{HF}_A}{H_A}{\nu_A}  \\
        \boldsymbol{0} & 0 & \delta_{\mu_B\nu_B}\Tbraket{\mu_A}{H_A}{\text{HF}_A} & \delta_{\mu_B\nu_B} H^{A}_{\mu_A\nu_A} + \delta_{\mu_A\nu_A} H^{B}_{\mu_B\nu_B}\\
        \end{pmatrix}\label{eq:simH}
\end{align}
\end{figure*}
We now investigate the size-extensivity of the RAS-CC framework. This can be evaluated by considering two non-interacting subsystems $A$ and $B$, and computing separately the energy of the complex, $E_{AB}$, and of the monomers,  $E_{A}$ and $E_{B}$. The energy difference,
\begin{align}
    \Delta E = E_{AB} - (E_A + E_B),
\end{align}
should then be zero for a size-extensive model. Coupled cluster theory is size-extensive at all levels of the theory, while truncated configuration interaction (and RAS-CI) is not. Here, we consider two cases: 1) RAS II is divided between subsystems $A$ and $B$ and 2) RAS II is located on subsystem $A$. 
To illustrate we consider the four systems shown in Figure \ref{fig:SEsystems}: \ce{2CH2}, \ce{2CH2} + \ce{2He}, \ce{2CH3}, and \ce{H2O}+\ce{2He}. 
In case 1, the complete active space of the complex should be the sum of the complete active spaces of the monomer calculations.
For instance, for \ce{2CH_3} the active space for the complex equals that of the ethylene dissociation, CAS(4,4), while the monomer calculations use a CAS(2,2).
In Table \ref{tab:SEnumbers}, we show that for this type of subsystem partitioning, the RAS-CCSD approach has significant size-extensivity errors (though the errors are significantly smaller than for RAS-CISD). Comparing \ce{2CH2} with \ce{2CH3}, we see, however, that the size-extensivity error does not scale with the total number of electrons, but with the size of the active space; for \ce{2CH3} the complex calculation uses a CAS(2,2).
For case 2, the RAS-CCSD approach is size-extensive, as is seen from \ce{H2O}+\ce{2He}, with a CAS(2,2) on \ce{H2O}, or in the extensivity error for the ethylene dissociation products when two additional \ce{He} atoms are included in the calculation. In the \textit{Supporting Information}$^\dag$, we demonstrate that for case 2 systems, also the excitation energies on the $A$ system are intensive. 
We emphasize that in the RAS-CCSD model, where the active space is always taken to be the minimal, type 2 size-extensivity is of higher importance for case 1.
Nevertheless, we note that the type 1 size-extensivity error can be reduced by e.g., allowing additional particles in RAS III. 

%In passing, we analyze the block structure of the similarity transformed Hamiltonian matrix in case 2. The Hamiltonian is written as $\simH = H_A + \simH_B$, as $T = T_B$. The basis is split in the following way: $ \bra{HF}$, $\bra{\mu_A}$, $\bra{\mu_B}$ and $\bra{\mu_{AB}}$. From the block structure in eq. (11), we observe that the $A$ block is decoupled from the $B$ and $AB$ blocks, and that the block is shifted by $E^B_{\text{CC}}$. This implies that the ground state energy, and excited state energies corresponding to excitations in $A$, are size-extensive.  In the case where the $T$ operator also include excitations among doubly occupied and virtual orbitals in $A$, the block structure remains the same.Further, we note that the the $B$ and $AB$

In passing, we analyze the block structure of the similarity transformed Hamiltonian matrix in case 2. The Hamiltonian is written as $\simH = H_A + \simH_B$, as $T = T_B$. The basis is split in the following way: $ \bra{HF}$, $\bra{\mu_A}$, $\bra{\mu_B}$ and $\bra{\mu_{AB}}$. From the block structure in eq. \eqref{eq:simH}, we observe that the $A$ block only coupled to the reference block, but is decoupled from the $B$ and $AB$ blocks. The block is shifted by $E^B_{\text{CC}}$. This implies that the ground state energy, and excited state energies corresponding to excitations in $A$, are size-extensive. 
In the \textit{Supporting Information}$^\dag$, we report numerical evidence that excited states in $A$ remain size-extensive when the $T$ operator also includes excitations among doubly occupied and virtual orbitals in $A$.
%The block structure remains the same in the case where the $T$ operator also include excitations among doubly occupied and virtual orbitals in $A$.
We also note that the $B$ block is coupled to the $AB$ block through $\Tbraket{\mu_{AB}}{\simH}{\nu_B} = \delta_{\mu_B\nu_B}\Tbraket{\mu_A}{H_A}{\text{HF}_A}$, and that the RAS-CCSD excitation energies of $B$ will not, in general, be equal to the EOM-CCSD excitation energies of $B$. However, they will be size-intensive with respect to increasing the size of $B$, the \textit{Supporting Information}$^\dag$.

\section*{Summary and concluding remarks}
In this paper, we have presented a systematically improvable and pragmatic, yet balanced multireference coupled cluster framework. We have shown that the theory avoids the non-convexity issues of standard coupled cluster theory close to ground state degeneracies, though the theory remains non-Hermitian and will give complex energies in, and close to, conical intersections. The model is simple both to formulate and implement. The RAS-CC models sidestep the problems (with non-commuting terms in the cluster operator) that other balanced multireference, or spin-adapted open-shell formulations of coupled cluster theory encounters. 
For the ground state, RAS-CCSD has similar accuracy as the closed-shell CCSD or UCCSD, where those models are applicable. However, the RAS-CCSD approach can be used in cases when the standard formulation of the theory breaks down. Examples are dissociation processes or for open-shell states where the reference wavefunction is not a single determinant. Finally, we demonstrate that for a fixed active space, the correlation energy has the correct scaling properties with system size. 

\section*{Author Contributions}
Sarai Dery Folkestad: Conceptualization (equal); Methodology; Software; Formal analysis; Investigation; Writing, original draft (lead); Visualization.
Henrik Koch: Conceptualization(equal); Writing, original draft (supporting)

\section*{Conflicts of interest}
The authors declare no conflict of interest.

\section*{Data availability}
Geometries, details of 2D scans and loop trajectories are available in the \textit{Supporting Information}.$^\dag$

\section*{Acknowledgments}
The authors thank Federico Rossi and Leo Stoll for helpful discussions.

\bibliography{rsc} %You need to replace "rsc" on this line with the name of your .bib file
\bibliographystyle{rsc} %the RSC's .bst file

\onecolumn



\section*{Supplementary material}

# S1 Ethylene: additional results

## Details of conical intersection $gh$-plane scan and loop.

Loop scan is made by displacements

$$\boldsymbol{R}(r,\theta) = \boldsymbol{R}_0 + r(\cos\theta\,\hat{\boldsymbol{g}} + \sin\theta\,\hat{\boldsymbol{h}}), \tag{S1}$$

with $r = 0.25$, and $\boldsymbol{R}_0$ defined in Table S1, and the vectors $\boldsymbol{g}$ and $\boldsymbol{h}$, defined in Tables S3 and S4, are normalized and orthogonalized to obtain $\hat{\boldsymbol{g}}$ and $\hat{\boldsymbol{h}}$. The 2D scans are performed with

$$\boldsymbol{R}(r,\theta) = \boldsymbol{R}_0 + \delta g \cdot \boldsymbol{g} + \delta h \cdot \boldsymbol{h}, \tag{S2}$$

with $\boldsymbol{R}_0$ geometries refined in Tables S1 and S2, and the vectors$\boldsymbol{g}$ and $\boldsymbol{h}$, defined in Tables S3 and S4, are normalized and orthogonalized to obtain $\hat{\boldsymbol{g}}$ and $\hat{\boldsymbol{h}}$. The calculation sequence is visualized in Figure S1.

Table S1: Reference geometry ($\boldsymbol{R}_0$) for the ethylene/cc-pVDZ $S_0/S_1$ conical intersection CCSD 2D-scan, and CCSD, CCSDT, RAS-CCSD, and RAS-CISD loop scans. Atomic units are used.

| Atom | $x$ | $y$ | $z$ |
|---|---|---|---|
| H | 1.997867955695 | -0.934072866019 | 1.400229198262 |
| C | 1.148640897418 | 0.239876842417 | -0.131392708244 |
| H | 2.560613477894 | 0.855538648059 | -1.549537590664 |
| C | -1.453179785883 | 0.709122220280 | -0.488417346057 |
| H | -2.628808023756 | 0.852446686508 | 1.268900103872 |
| H | -1.625277793952 | -1.533961242001 | -0.500516823231 |

Table S2: Reference geometry ($\boldsymbol{R}_0$) for the ethylene/cc-pVDZ $S_0/S_1$ conical intersection RAS-CCSD 2D-scan. Atomic units are used.

| Atom | $x$ | $y$ | $z$ |
|---|---|---|---|
| H | 1.999248707990 | -0.932654648943 | 1.401572255919 |
| C | 1.141402436251 | 0.236048276704 | -0.129450513335 |
| H | 2.560435308712 | 0.854309269033 | -1.550603795346 |
| C | -1.451701995448 | 0.717363155912 | -0.492667119115 |
| H | -2.627223544648 | 0.848946167203 | 1.271174533659 |
| H | -1.622267231975 | -1.535079970434 | -0.500743769949 |

Table S3: The ***g***-vector used in the ethylene/cc-pVDZ $S_0/S_1$ conical intersection 2D- and loop scans. Atomic units are used.

| Atom | $x$ | $y$ | $z$ |
|---|---|---|---|
| H | 0.007313466289 | 0.057147259497 | 0.037233885635 |
| C | 0.018065045039 | 0.010653044239 | 0.029770694182 |
| H | -0.011045778402 | -0.044246241553 | -0.025640855466 |
| C | -0.016011637809 | -0.042775821534 | 0.029986440784 |
| H | 0.030212324793 | -0.002500610868 | 0.030496622594 |
| H | -0.028533419910 | 0.021722370219 | -0.101846787729 |

Table S4: The ***h***-vector used in the ethylene/cc-pVDZ $S_0/S_1$ conical intersection 2D- and loop scans. Atomic units are used.

| Atom | $x$ | $y$ | $z$ |
|---|---|---|---|
| H | 0.009983729150 | 0.004279864574 | 0.006085526830 |
| C | -0.059128405254 | -0.031406366244 | 0.011698042309 |
| H | -0.000072261588 | -0.004346978757 | -0.005302673058 |
| C | 0.013554961398 | 0.069990614294 | -0.037047741578 |
| H | 0.008830292504 | -0.027241876753 | 0.014224759752 |
| H | 0.027122442072 | -0.011417198126 | 0.010473940680 |

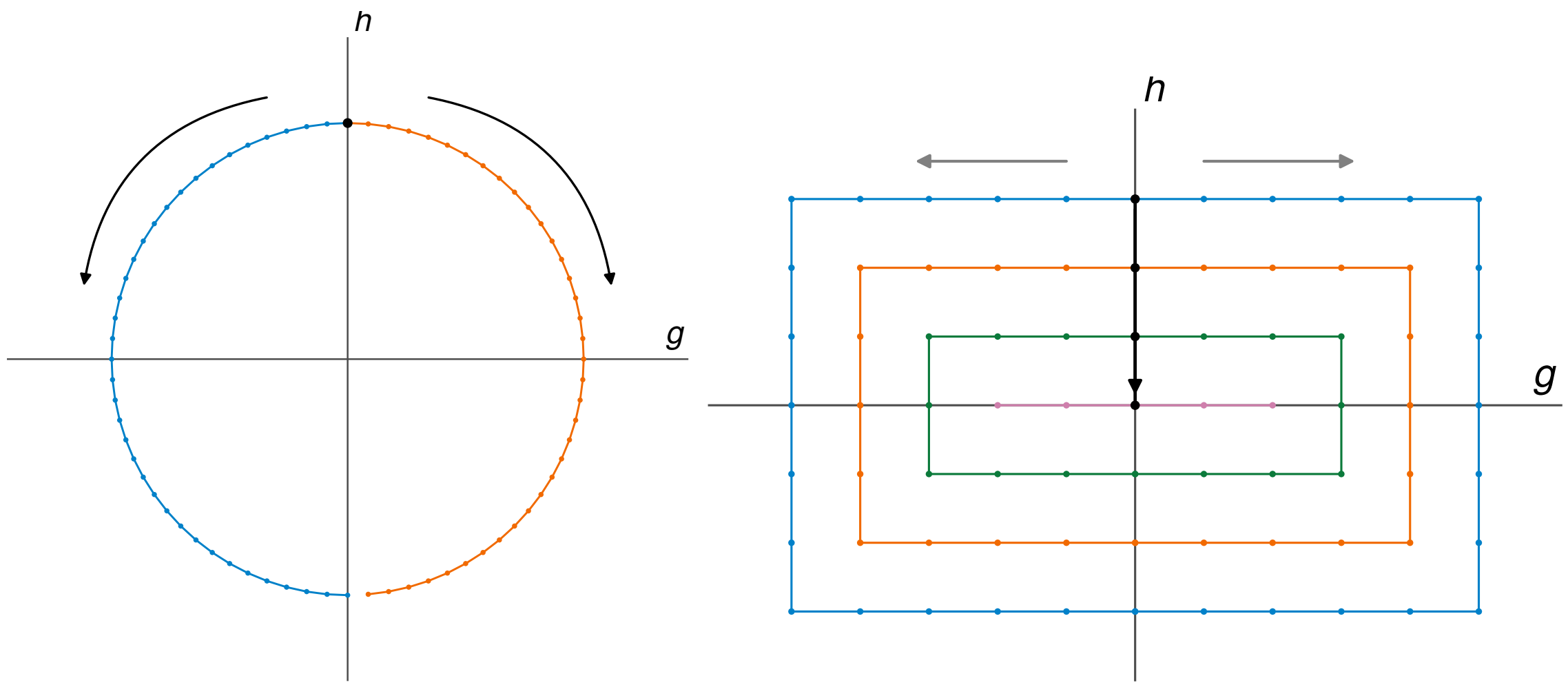


Figure S1: The scanning sequence for the conical intersection loop (left) and 2D (right) scans.

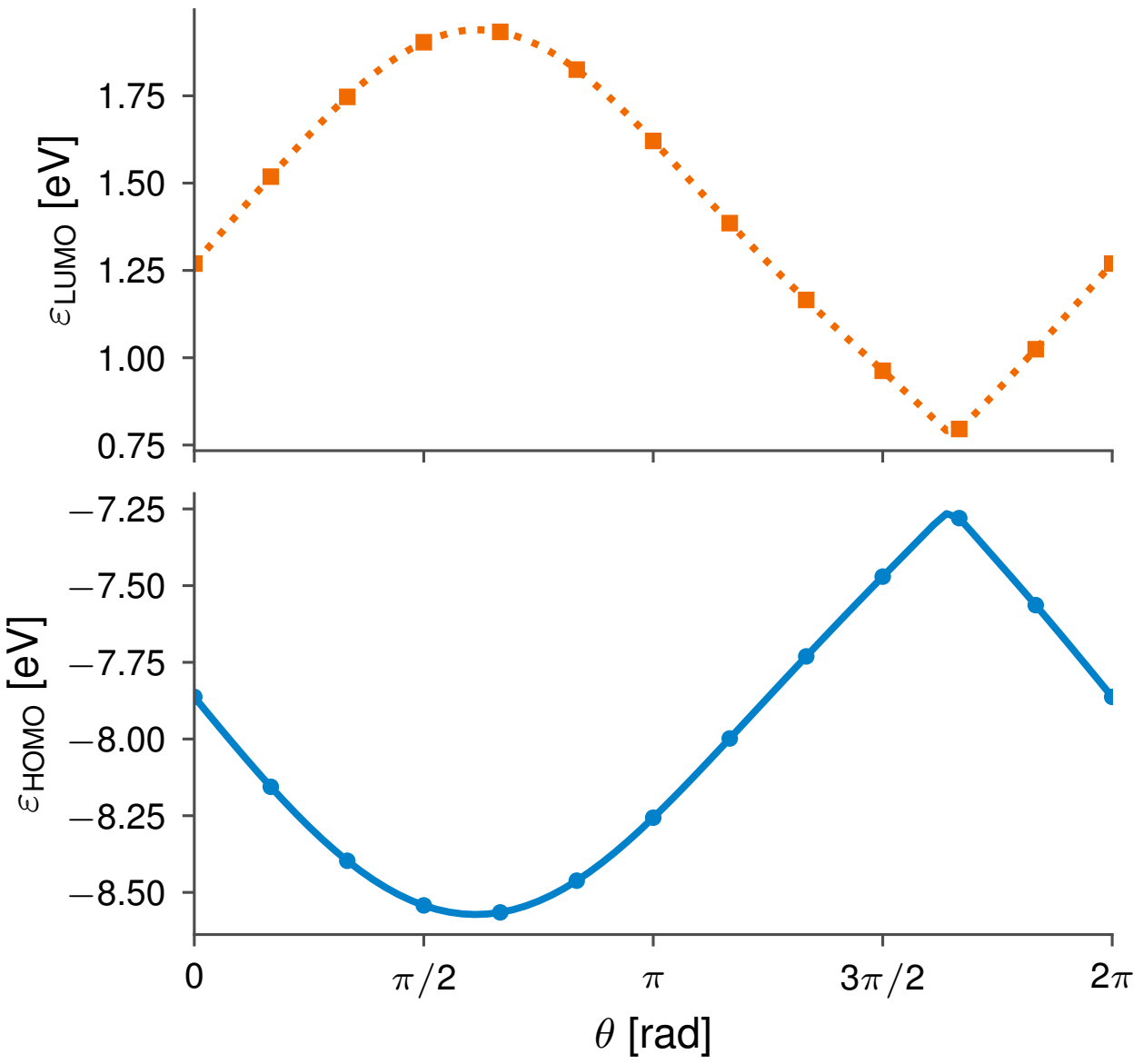


Figure S2: Ethylene/cc-pVDZ restricted Hartree–Fock LUMO (top) HOMO (bottom) orbital energies for a $r = 0.25\,a_0$ loop in the $gh$-plane (defined in Tables S3 and S4) around the point in Table S1.

### Torsion and stretch scans

The reference geometry is the planar ethylene ($D_{2h}$) with the C–C bond along $x$, see Table S5. The torsion rotates one of the $CH_2$ group about the C=C axis by $\theta$, with the $CH_2$ at fixed bond bond lengths, $\theta = 0$ defines the planar geometry and $\theta = \frac{\pi}{2}$ the twisted $D_{2d}$ geometry.

Table S5: Reference geometry of ethylene shared by the torsion and stretch scans in Å.

| Atom | $x$ | $y$ | $z$ |
|---|---|---|---|
| C | 0.6669 | 0.0000 | 0.0000 |
| C | −0.6669 | 0.0000 | 0.0000 |
| H | 1.2295 | 0.9223 | 0.0000 |
| H | 1.2295 | −0.9223 | 0.0000 |
| H | −1.2295 | 0.9223 | 0.0000 |
| H | −1.2295 | −0.9223 | 0.0000 |

In Figure S3, we show the ethylene/cc-pVDZ torsion using a restricted Hartree–Fock reference for all calculations, including for the RAS-based methods.

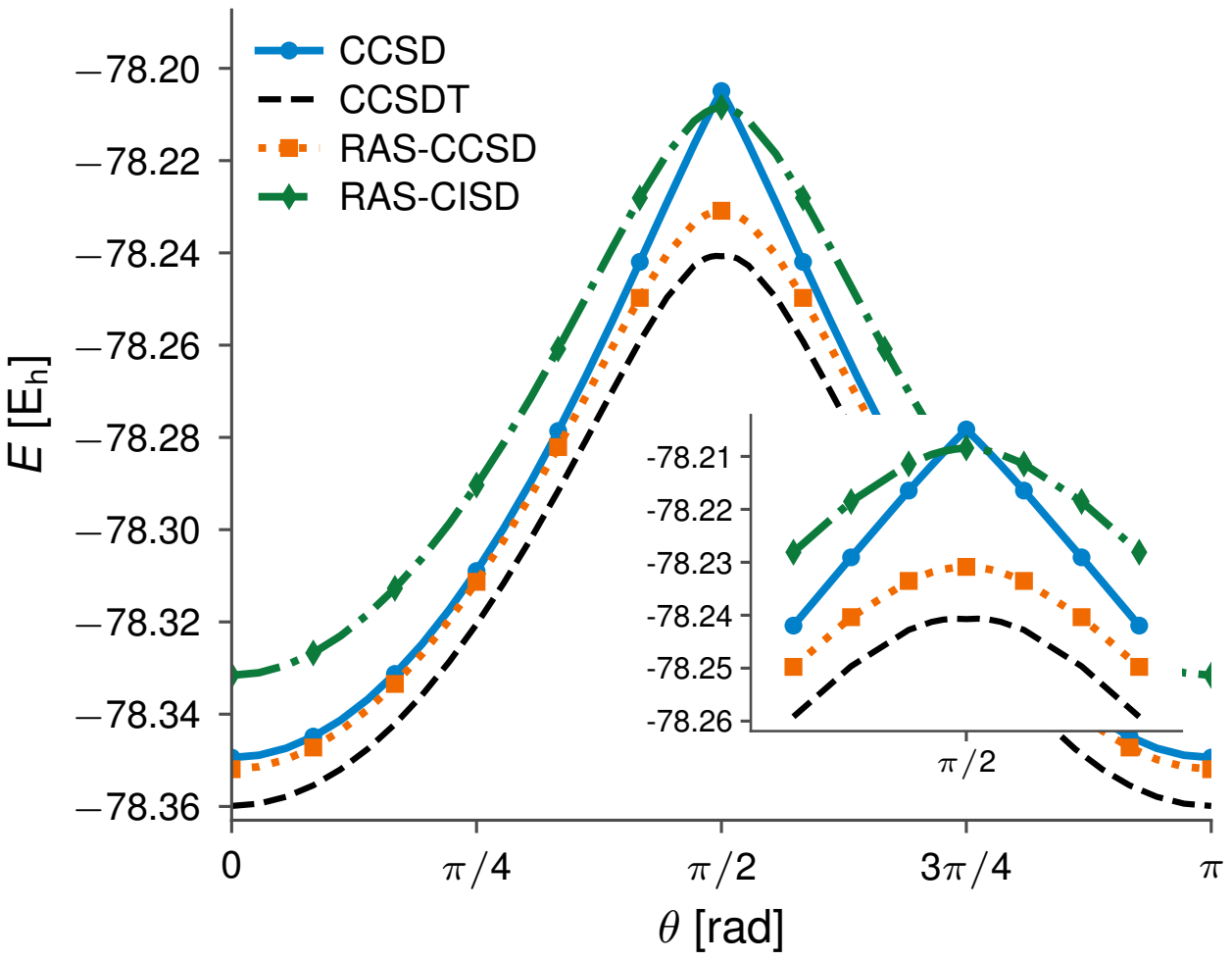


Figure S3: Ethylene/cc-pVDZ C-C bond torsion (unrelaxed) with CCSD, CCSDT, RAS-CCSD and RAS-CISD. A restricted Hartree–Fock reference is used in all calculations.

## S2 Triple bond dissociation

The C–N bond lies along $x$. The scan displaces the N along $+x$ with the $CH_3$–C fragment held fixed. The scan coordinate $R$ is the C–N distance and the N is located at $x = 0.1006 + R$.

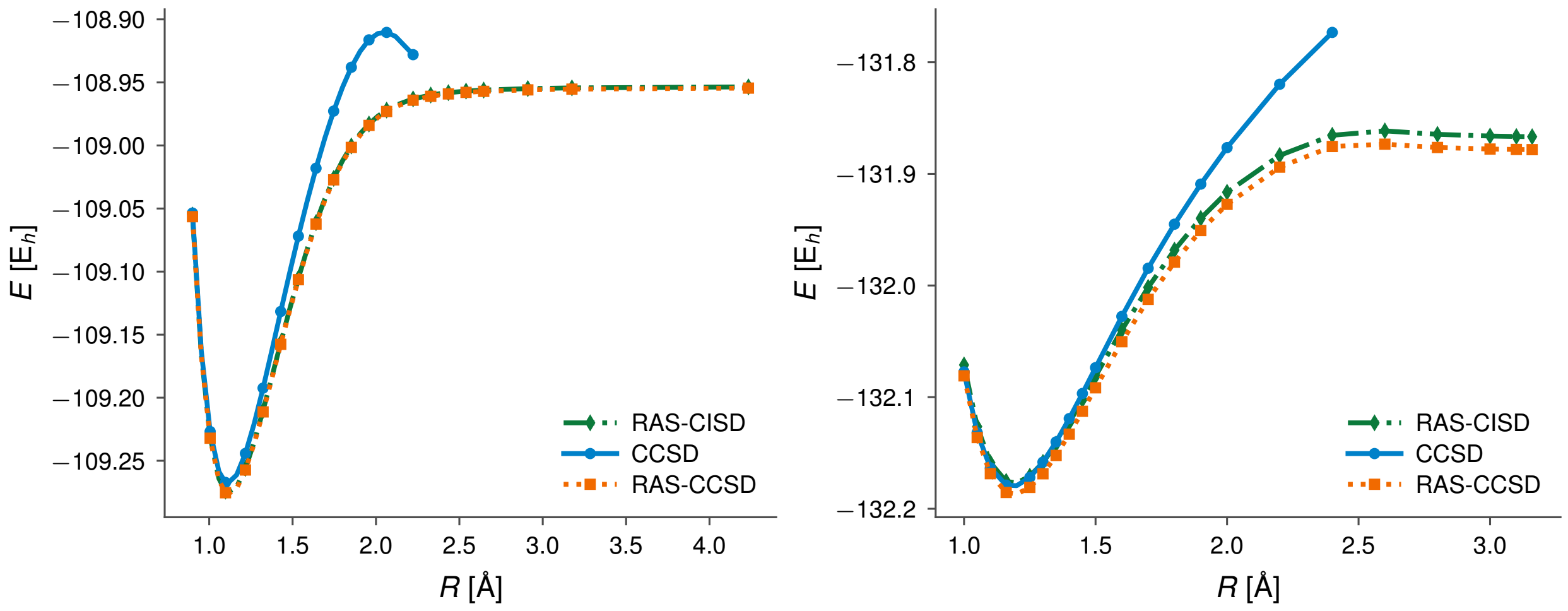


Figure S4: $N_2$/cc-pVDZ dissociation (left) and the unrelaxed C–N dissociation in $CH_3CN$/6-31G (right) calculated with the CCSD, RAS-CISD, and RAS-CCSD approaches.

Table S6: Reference geometry for $CH_3CN$ in Å (for $R = 1.1602$ Å).

| Atom | $x$ | $y$ | $z$ |
|---|---|---|---|
| N | 1.2608 | 0.0000 | 0.0000 |
| C | 0.1006 | 0.0000 | 0.0000 |
| C | −1.3613 | 0.0000 | 0.0000 |
| H | −1.7500 | −0.8301 | 0.5974 |
| H | −1.7501 | −0.1022 | −1.0175 |
| H | −1.7500 | 0.9324 | 0.4202 |

## S3 Additional numerical size-extensivity tests

In Table S7, we show the the size-extensivity error of the ten lowest RAS states of $H_2O$ with two He atoms, each 200 Å from the oxygen, $\Delta E = E(H_2O + 2\,He) - [E(H_2O) + 2E(He)]$. A CAS(2,2) active space is located on the water molecule. All energies are given in $E_h$. The RAS-CISD approach is seen to have significant errors for all states, whereas RAS-CCSD is size-extensive for all states.

Table S7: Size-extensivity of the ten lowest RAS states of $H_2O$ with two He atoms, each 200 Å from the oxygen, using the cc-pVDZ basis. $\Delta E = E(H_2O + 2\,He) - [E(H_2O) + 2E(He)]$. A CAS(2,2) active space is located on the water molecule. All energies are given in $E_h$.

| | RAS-CCSD | | | RAS-CISD | | |
|---|---|---|---|---|---|---|
| State | Complex | Fragments | $\Delta E$ | Complex | Fragments | $\Delta E$ |
| 1 | −82.012074647953 | −82.012074647955 | $2.00 \times 10^{-12}$ | −81.999063773888 | −82.005321375787 | $6.26 \times 10^{-3}$ |
| 2 | −81.739738064208 | −81.739738064207 | $-9.95 \times 10^{-13}$ | −81.718897984759 | −81.727751005511 | $8.85 \times 10^{-3}$ |
| 3 | −81.714382285845 | −81.714382285845 | 0 | −81.694050598623 | −81.702763025983 | $8.71 \times 10^{-3}$ |
| 4 | −81.659689578327 | −81.659689578327 | 0 | −81.533612489807 | −81.598481197504 | $6.49 \times 10^{-2}$ |
| 5 | −81.635693359414 | −81.635693359414 | 0 | −81.500375552125 | −81.565232122929 | $6.49 \times 10^{-2}$ |
| 6 | −81.625894602878 | −81.625894602877 | $-1.01 \times 10^{-12}$ | −81.480387891867 | −81.545256599564 | $6.49 \times 10^{-2}$ |
| 7 | −81.621622199141 | −81.621622199140 | $-9.95 \times 10^{-13}$ | −81.466225122606 | −81.531093830302 | $6.49 \times 10^{-2}$ |
| 8 | −81.529379038170 | −81.529379038171 | $9.95 \times 10^{-13}$ | −81.407347626757 | −81.472216334454 | $6.49 \times 10^{-2}$ |
| 9 | −81.497609745784 | −81.497609745784 | 0 | −81.376097193769 | −81.440965901465 | $6.49 \times 10^{-2}$ |
| 10 | −81.489778824562 | −81.489778824563 | $1.01 \times 10^{-12}$ | −81.370923814411 | −81.435792522106 | $6.49 \times 10^{-2}$ |

In Table S8, we consider the case of excitations in $B$, following the discussion around eq. (11). We show that while the excitation energies of $B$ are not the same as in EOM-CCSD, they are size extensive with respect to an increase of $B$.

Table S8: The lowest singlet and triplet excitation energies of $H_2O$ in $A + B$ complexes at 200 Å separation, with the CAS(2,4) active space placed on Be ($A$). Adding a He atom, also at 200 Å, leaves both excitation energies unchanged. The RAS-CCSD excitation energies of $B$ differ from EOM-CCSD, but they are size extensive with respect to an increase in $B$.

| System | $\omega_T^{H_2O} [E_h]$ | $\omega_S^{H_2O}$ $[E_h]$ |
|---|---|---|
| $H_2O$ + Be | 0.275659733526 | 0.301591903119 |
| $H_2O$ + Be + He | 0.275659733527 | 0.301591903120 |
| Difference | $1.0 \times 10^{-12}$ | $1.0 \times 10^{-12}$ |
| $H_2O$, EOM-CCSD | 0.267698282893 | 0.293331402780 |
| Deviation from EOM-CCSD [m$E_h$] | 7.96 | 8.26 |



\end{document}